\documentstyle[prl,aps,psfig,twocolumn,floats]{revtex}
\begin{document}
\draft
\wideabs{
\title{Observation of the parallel-magnetic-field-induced
superconductor-insulator transition in thin amorphous InO films}
\author{V.F.~Gantmakher\thanks{e-mail: gantm@issp.ac.ru},
M.V.~Golubkov, V.T.~Dolgopolov, A.A.~Shashkin, and G.E.~Tsydynzhapov}
\address{Institute of Solid State Physics RAS, Chernogolovka, Moscow
District 142432, Russia}
\maketitle
\begin{abstract}
We study the response of a thin superconducting amorphous InO film
with variable oxygen content to a parallel magnetic field. A
field-induced superconductor-insulator transition (SIT) is observed
that is very similar to the one in normal magnetic fields. As the
boson-vortex duality, which is the key-stone of the theory of the
field-induced SIT, is obviously absent in the parallel configuration,
we have to draw conclusion about the theory insufficiency.
\end{abstract}
}

A good deal of work was performed on the investigation of the
magnetic-field-induced superconductor-insulator transition (SIT) on
superconducting amorphous films of InO \cite{Paa,gant,gant1}, MoGe
\cite{Kapit}, and MoSi \cite{Okuma} with thickness comparable to the
superconducting coherence length. It was found that with increasing
field $B$ the resistance $R$ of all studied films rises abruptly at a
magnetic field $B_c$ and then passes through a maximum followed by a
drop in high magnetic fields \cite{gant,gant1}. The film state just
above $B_c$ was identified as insulating although at lowest
temperatures about 30~mK the maximum resistance does not exceed
100~k$\Omega$ and the temperature dependences of $1/R$ correspond to
the activation energies which do not exceed by far the lowest
temperatures. Near $B_c$ at sufficiently low temperatures, the
resistance $R(T,B)$ was found to be a function of single scaling
variable $u=(B-B_c)/T^y$ with exponent $y\approx 0.8$. The above
experimental findings are regarded to confirm the theory of the
quantum SIT \cite{th1} in two-dimensional (2D) superconducting films
subjected to a normal magnetic field. This theory exploits the
concept of the hypothetical system of charged bosons in a random
potential and is based on the boson-vortex symmetry of the model
Hamiltonian. In the vicinity of the SIT point ($T=0,B=B_c$) the film
resistance $R(T,B)$ is expected to be a universal function of the
single scaling variable which is defined as the ratio of the
correlation length $\xi\propto (B-B_c)^{-\nu}$ and the dephasing
length $L_\phi\propto T^{-1/z}$, where $\nu$ and $z$ are the critical
indices. The form of the scaling variable implies that the value
$1/z\nu$ has to be identified with exponent $y$. The concept of the
localization of electron pairs, or bosons \cite{th1}, has been
supported recently by the work of Ref.~\cite{Larkin}. There, it is
shown that for a 2D superconducting film with strong disorder the
region of fluctuation superconductivity, where the electron pairs
occur, should extend down to zero temperature. In this region the
unpaired electrons are supposed to be localized whereas the bosons
can be either localized or delocalized, dependent on the value of
magnetic field.

Other experimental results were interpreted within the model of
electron pair localization: (i) a crossing of Hall isotherms
$R_{xy}(B)$ was observed at a field $B_{c0}>B_c$ and attributed to a
transition between the Bose-insulator and a Fermi-insulator that
consists of localized single electrons, i.e., pairing is presumed to
be destroyed at $B_{c0}$ \cite{Paa}; (ii) the resistance drop in high
fields was explained in terms of the electron pair breaking which
occurs gradually with increasing $B$ owing to the different binding
energies of bosons in a random potential \cite{gant,gant1}.

However, a very similar SIT has been observed recently on amorphous
MoSi films with the thickness $d=1700$~\AA\ which is an order of
magnitude larger compared to the superconducting coherence length
$\xi_{sc}$ \cite{Samoi}. This fact causes one to think that either
the theory is inadequate or its restrictions are too severe.

Here, we investigate the influence of a parallel magnetic field on
the superconducting properties of a thin amorphous InO film with
variable oxygen content. For all film states we find a complete
similarity in the behaviour of the resistance $R(T,B)$ for both
parallel and perpendicular magnetic field.

The sample is an amorphous InO film with thickness 200~\AA\ that was
grown by electron-gun-evaporation of a high-purity In$_2$O$_3$ target
onto a glass substrate \cite{thankyou}. Oxygen deficiency compared to
fully stoichiometric insulating compound In$_2$O$_3$ causes the film
conductivity. By changing the oxygen content one can cover the range
from a superconducting material ($\xi_{sc}\alt 500$~\AA ) to an
insulator with activated conductance \cite{ShOv}. The procedures to
change reversibly the film state are described in detail in
Ref.~\cite{gant}. To reinforce the superconducting properties of the
film we used heating in vacuum up to a temperature from the interval
70 -- 110$^\circ$C until the sample resistance got saturated. To
shift the film state in the opposite direction we made exposure to
air at room temperature. As the film remains amorphous during these
manipulations, it is natural to assume that the treatment used
results mainly in a change of the carrier density $n$ and that the
value $n$ is inversely proportional to the room temperature
resistance $R_r$ of the sample. Seven superconducting states of the
film were studied.

The film was mounted into the top loading system of a dilution
refrigerator; it was set either parallel or normal to the magnetic
field within $1^\circ$ accuracy. A four-terminal lock-in technique at
a frequency of 10~Hz was used to measure the resistance of the
sample. The ac current through the sample was equal to 1~nA.

\begin{figure}
\psfig{file=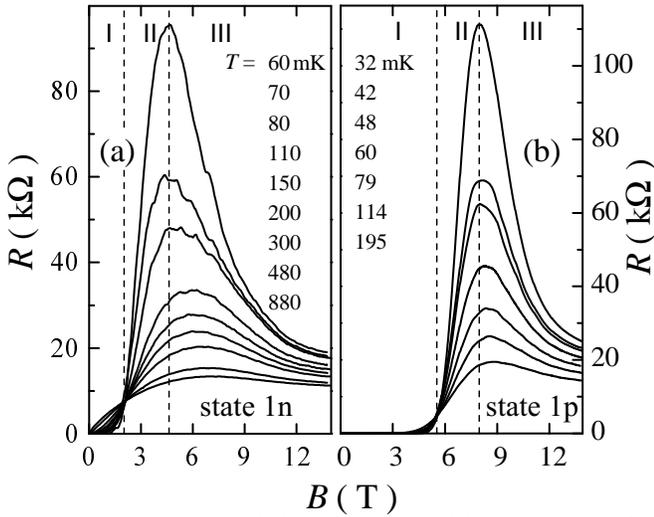,width=\columnwidth,clip=}
\caption{Isotherms $R(B)$ for a normal (a) and parallel (b) magnetic
field. The dashed lines separate regions I, II, and III and
correspond to the critical field $B_c$ and the resistance maximum at
the lowest temperatures.}
\label{f1}
\end{figure}

\begin{figure}[tb]
\psfig{file=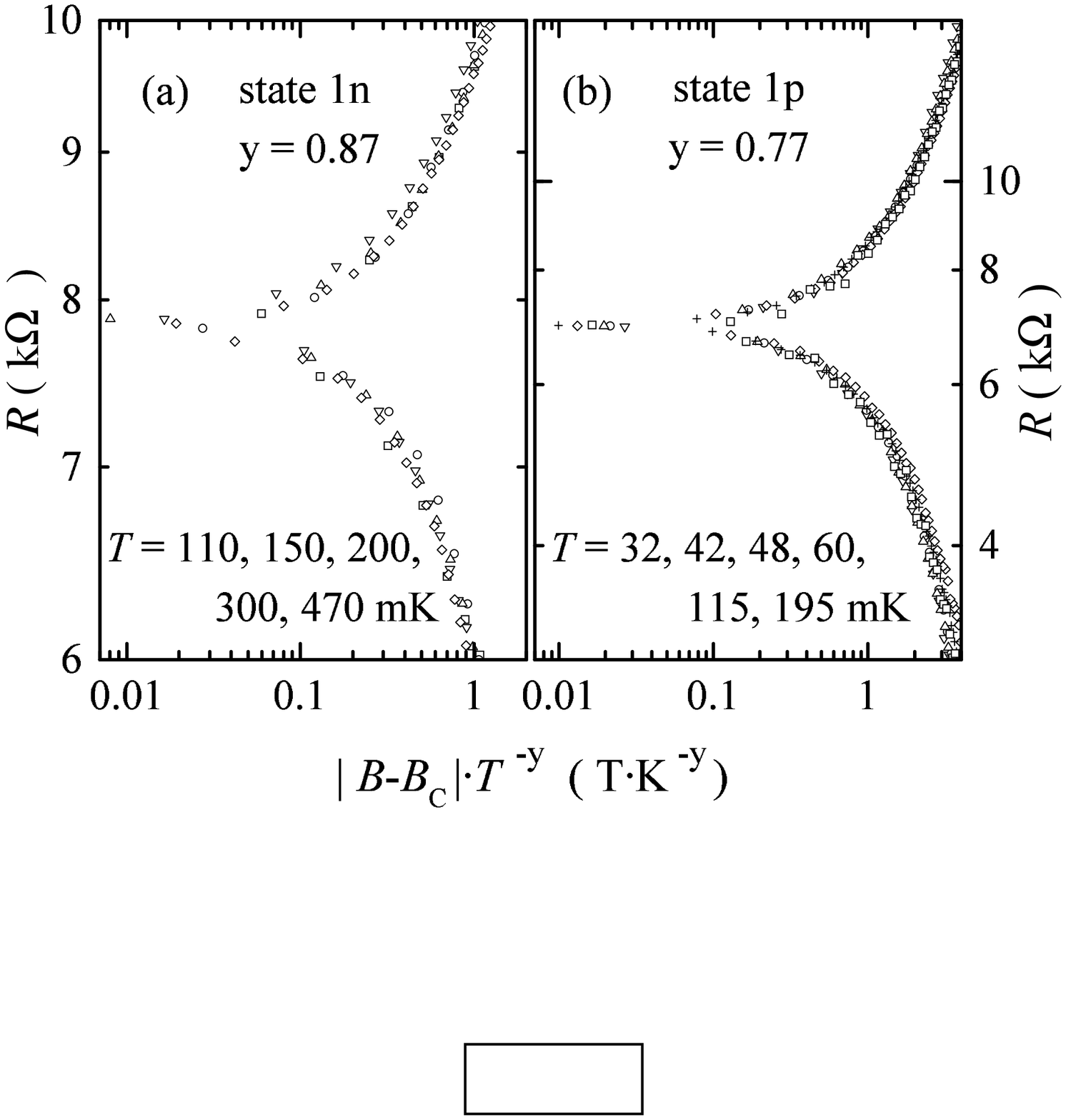,width=\columnwidth,clip=}
\caption{Scaling plot for both normal (a) and parallel (b) magnetic
field.}
\label{f2}
\end{figure}

The experimental dependences $R(B)$ for two close states of the film
in the normal and parallel field orientation are presented in
Fig.~\ref{f1}. As seen from the figure, for both field orientations
the isotherms cross at the critical field $B_c$ which separates the
superconducting region I and the insulating region II. The resistance
drop with $B$ in region III down to values about $h/e^2$ indicates
that at high magnetic fields the film state is metallic. The critical
field $B_c$ is found to vary with field direction by about a factor
of two. Nevertheless, near $B_c$, the experimental data collapse onto
a single curve equally well for both perpendicular and parallel
magnetic field with the close values of exponent $y$, see
Fig.~\ref{f2}.

\begin{figure}[tb]
\centerline{\psfig{file=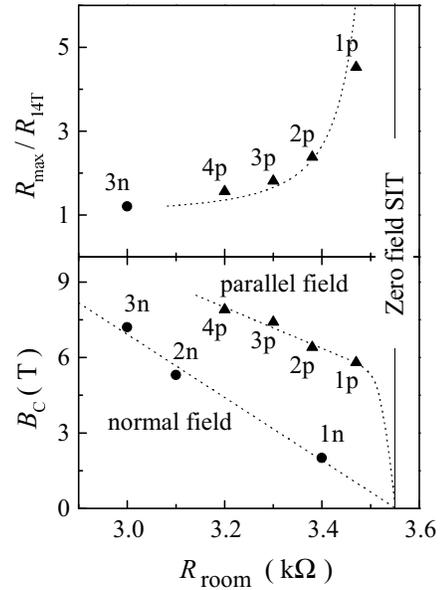,width=0.65\columnwidth,clip=}}
\caption{Change of the relative resistance maximum at $T\approx
30$~mK and of the critical magnetic field with changing film state as
determined by the room temperarure resistance of the film. The dashed lines
are guides to the eye.} \label{f3}
\end{figure}

Figure~\ref{f3} displays the behaviour of the relative resistance
maximum $R_{\rm max}/R_{\rm 14T}$ at the lowest temperature and of
the critical field $B_c$ with changing film state. As seen from the
figure, the resistance ratio decreases and approaches unity as one
goes deeper into the superconducting phase. The value $B_c$ for both
field directions increases when departing from the zero-field SIT,
being higher in the parallel configuration.

\begin{figure}[tb]
\psfig{file=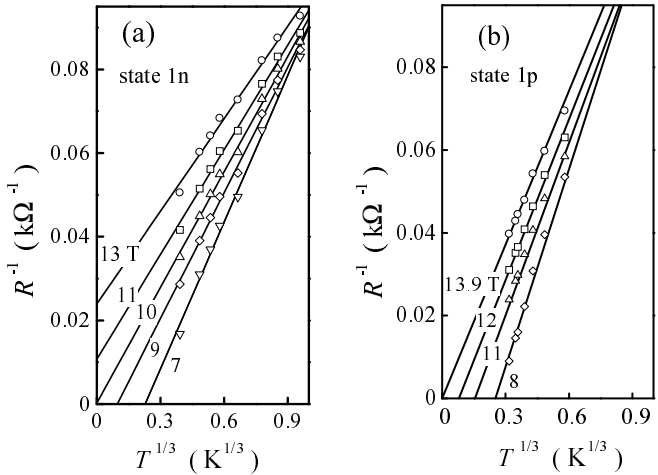,width=\columnwidth,clip=}
\caption{The temperature dependence of $1/R$ near the
three-dimensional metal-insulator transition in a normal (a) and
parallel (b) magnetic field.}
\label{f4}
\end{figure}

The mean free path of the normal electrons in our film is small
compared to the film thickness \cite{gant1} and, hence, the
metal-insulator transition expected in region III should be
three-dimensional. In this case the conductance $1/R$ in the vicinity
of metal-insulator transition is expected to change linearly with
$T^{1/3}$ and extrapolation of the linear dependence to $T=0$ should
reveal whether the phase is metallic or insulating as judged by
offset sign \cite{3d}. Such a data analysis is shown in
Fig.~\ref{f4}. For both field configurations the experimental data
behave similarly: the offset of the linear dependence increases with
$B$ passing through zero at the transition point.

The crucial point of the theory of the magnetic-field-driven SIT
\cite{th1} is the boson-vortex duality for induced by external
magnetic field vortices penetrating the film, which is not the case
for a parallel magnetic field. The experimentally observed complete
similarity of the SIT properties in the perpendicular and parallel
magnetic field forces one to conclude that the theory is not directly
applicable for actual superconducting systems with disorder.

Nonetheless, assuming that the parallel magnetic field is capable of
localizing the fluctuation-induced Cooper pairs in the
paraconductivity region, one can translate all experimental findings
into the language of localized bosons. We emphasize that the
speculations below are very attractive but do not have a sound
experimental basis so far.

One can reckon that the resistance rise with $B$ in region II is
caused by decreasing boson localization length. Assuming additionally
that the magnetic field not only localizes but also breaks electron
pairs, it is easy to interpret the resistance drop in region III:
breaking the correlations in localized electron pairs results in an
increase of the electron hopping probability \cite{gant} and
eventually electron delocalization at high fields \cite{gant1}. For
sufficiently deep film states in the superconducting phase, i.e.,
sufficiently high $B_c$, at $B>B_c$ the localized bosons coexist with
delocalized normal electrons so that the state never becomes
insulating. In other words, as one advances into the superconducting
phase, a fraction of the localized bosons reduces. As a result, the
relative amplitude of the resistance maximum $R_{\rm max}/R_{\rm
14T}$ tends to unity (Fig.~\ref{f3}), and the SIT should transform
into an ordinary superconductor to normal metal transition.

In summary, we have investigated the response of a thin
superconducting amorphous InO film to a parallel magnetic field. At a
critical field $B_c$ a SIT has been observed that is very similar to
the one in normal magnetic fields. That the boson-vortex duality is
absent in the parallel configuration points to the insufficiency of
the theory of the field-induced SIT \cite{th1}. We find that the
behaviours of the film resistance at fields above the transition are
also similar for the parallel and normal magnetic field.

This work was supported by Grants RFBR 99-02-16117 and RFBR-PICS
98-02-22037 and by the Programme "Statistical Physics" from the
Russian Ministry of Sciences.

\end{document}